\documentclass{acm_proc_article-sp}
\usepackage{makeidx}  % allows for indexgeneration

\usepackage{graphicx}
\usepackage{epstopdf}
\usepackage{comment}
\usepackage{makecell}
\usepackage[hyphens]{url}
\usepackage{booktabs}
\usepackage[usenames,dvipsnames,svgnames,table]{xcolor}
\usepackage{psfrag}
\usepackage{amsmath}
\newcommand{\ra}[1]{\renewcommand{\arraystretch}{#1}}

\newcommand{\ourname}{EldeRan}

\begin{document}

\title{Automated Dynamic Analysis of Ransomware: \\ Benefits, Limitations and use for Detection}
\numberofauthors{4} %  in this sample file, there are a *total*
% of EIGHT authors. SIX appear on the 'first-page' (for formatting
% reasons) and the remaining two appear in the \additionalauthors section.
%
\author{
% You can go ahead and credit any number of authors here,
% e.g. one 'row of three' or two rows (consisting of one row of three
% and a second row of one, two or three).
%
% The command \alignauthor (no curly braces needed) should
% precede each author name, affiliation/snail-mail address and
% e-mail address. Additionally, tag each line of
% affiliation/address with \affaddr, and tag the
% e-mail address with \email.
%
% 1st. author
%\alignauthor
Daniele Sgandurra, Luis Mu\~{n}oz-Gonz\'{a}lez,  Rabih Mohsen, Emil C. Lupu\\
       \affaddr{Department of Computing, Imperial College London}\\
       \affaddr{180 Queen's Gate, SW7 2AZ, London, UK}\\
       \email{\{d.sgandurra, l.munoz, r.mohsen11, e.c.lupu\}@imperial.ac.uk}
}

%\date{30 September 2015}
% Just remember to make sure that the TOTAL number of authors
% is the number that will appear on the first page PLUS the
% number that will appear in the \additionalauthors section.

\maketitle

\hyphenation{ana-ly-sing}

\begin{abstract}
Recent statistics show that in 2015 more than 140 millions new malware samples have been found. Among these, a large portion is due to ransomware, the class of malware whose specific goal is to render the victim's system unusable, in particular by encrypting important files, and then ask the user to pay a ransom to revert the damage. Several ransomware include sophisticated packing techniques, and are hence difficult to statically analyse.  We present \emph{\ourname{}}, a machine learning approach for dynamically analysing and classifying ransomware. \ourname{} monitors a set of actions performed by applications in their first phases of installation checking for characteristics signs of ransomware. Our tests over a dataset of $582$ ransomware belonging to $11$ families, and with $942$ goodware applications, show that \ourname{} achieves an area under the ROC curve of $0.995$. Furthermore, \ourname{} works without requiring that an entire ransomware family is available beforehand. These results suggest that dynamic analysis can support ransomware detection, since ransomware samples exhibit a set of characteristic features at run-time that are common across families, and that helps the early detection of new variants. We also outline some limitations of dynamic analysis for ransomware and propose possible solutions.
\end{abstract}



%----INTRODUCTION-----%
\section{Introduction} \label{sec:introduction}
In recent months, numerous Internet users have been victims of cyber-attack campaigns based on extortion mechanisms \cite{talos_ransomware}, \cite{chimera}. 
 \emph{Ransomware} has emerged as one of the most difficult scareware\footnote{A class of malware that use tricks to scare users, e.g. by displaying a warning purportedly from a law enforcement agency claiming users have downloaded illegal content.} to defend from, as it might be computationally infeasible to revert ransomware's damage \cite{teslacrypt_cisco}. There are two basic types of ransomware available in the wild: the first one, \emph{locker-ransomware}, is designed to lock the victims' computer, to prevent them from using it; the second one, and most common nowadays, is \emph{crypto-ransomware}, which encrypts personal files to make them inaccessible to its victims. In both cases, users are forced to pay a ransom to regain access either to their data (assuming no backup mechanism is in place) or system. Ransomware has featured prominently in the media, in particular when in 2014 a Police Department's computer system was infected by \emph{Cryptowall} \cite{seacoastonline}.

Many victims feel their data are so important that they have to pay the ransom. For example, when in 2012 Symantec was able to dismantle a Command and Control (C\&C) network used by the \emph{CryptoDefense} ransomware family, the follow-on study showed that 2.9\% of victims, out of 68,000 unique infections, appeared to have paid the ransom. Extortion mechanisms, therefore, generate significant revenue for the attackers. For \emph{CryptoWall} version 3, statistics account for an estimated total \$325 million in damages in the US alone \cite{CTA_ransomware}. Note that the average ransom request amounts to \$300~\cite{symantec_growing}, and the preferred payment method is through Bitcoins, due to their untraceable properties. Anti-virus (AV) vendors strive to keep the pace with sophisticated malware variants. In fact, signature-based mechanisms are prone to being easily deceived, especially when new variants are spread. For example, according to \cite{talos_ransomware}, out of the 3,000 unique analysed exploit kits, which are one of the main vectors of ransomware (e.g., around 62\% of infections of Angler exploit kit deliver ransomware), only 6\% of their signatures were found in VirusTotal. Furthermore, the average detection of these signatures was very low, with usually less than ten engines (out of around 60) able to detect them. Furthermore, current ransomware implement very sophisticated packing techniques to evade detection, e.g. obfuscated API calls \cite{evade_detection_packers}, rendering the static analysis (e.g., of import tables) useless. As signature-based AVs are easily evaded, e.g by the appearance of new variants, it is fundamental to exploit alternative approaches based on dynamically analysing behaviours. In particular, machine learning has been widely applied in research to classify malware, as an alternative to signature-based approaches \cite{graph_based}, \cite{kolter2006learning}, \cite{Rieck:2011:AAM:2011216.2011217}.

In this paper, we propose \emph{\ourname{}}, a machine learning approach to classify ransomware based on their early actions. The underling assumptions of this work are based on the observation that ransomware contain unique dynamic features and, to stop their spread, it is crucial to detect new variants during their first appearance. To this end, \ourname{} firstly selects the relevant features that characterize the ransomware behaviour, and then classify each newly-installed application on a user PC through a machine learning algorithm as to perform detection without relying on classical heuristic or signature-based techniques. The goal of this work is to understand whether ransomware can be identified, with a high degree of accuracy, using a limited number of characteristic features and before infecting victims. Furthermore, \ourname{} provides an automatic way for creating signatures for new variants of ransomware families. \ourname{} complements well AVs signature-based mechanisms, as it can be used to identify cases where AVs may have missed new, or unknown, ransomware families. The main contributions of this paper are:

\begin{itemize}
\item We propose \ourname{}, a framework to identify the most significant ransomware dynamic features, and use them to detect ransomware. Through the Mutual Information criterion, we have identified the most relevant dynamic features amongst a large set of considered ones. \ourname{} exploits a relatively small set of features without reducing the performance of the machine learning classifier. By following this approach, \ourname{} is also well suited to detect new ransomware families. 
\item We have evaluated the accuracy of the Regularized Logistic Regression by comparing it against other machine learning classifiers, namely the Support Vector Machine (SVM) and Naive Bayes. We found that the Logistic Regression outperforms Naive Bayes and it is competitive with respect to the SVM. Moreover, the Logistic Regression is easy to train and adapt compared to the SVM. The regularization technique applied to the Logistic Regression helps the classifier to generalize better to unseen samples by preventing overfitting. 
\item We compare the classification results with those of VirusTotal: \ourname{}'s average error rate is 2.4\% while that of VirusTotal is 5.6\%, and \ourname{} achieves a remarkable 96.3\% detection rate. We also tested the ability of \ourname{} to detect new families of ransomware, obtaining an average detection rate of 93.3\%. 
\end{itemize}

The rest of the paper is organized as follows. 
In Section~\ref{sec:ransomware} we give an overview of ransomware.
Sect.~\ref{sec:analysis} details our approach (\emph{\ourname{}}) based on analysing a set of features in a sandboxed environment to detect variants of ransomware. In Sect.~\ref{sec:dataset} we analyse the most relevant features exhibited by the samples of our dataset based on the \ourname{}'s feature selection algorithm.  The results of the experimental evaluation is described in Sect.~\ref{sec:experiments}. In Sect.~\ref{sec:discussion} we discuss the relevant findings of our paper, the limitations of dynamic analysis, and suggestions of improvements. Finally, Sect.~\ref{sec:related} discusses related work, while in Sect.~\ref{sec:conclusion} we conclude the paper.

%---RANSOMWARE_SEC----%

\section{Ransomware} \label{sec:ransomware}

\emph{Ransomware} is any class of malware whose specific goal is to force users to pay a ransom to be able to regain full access to their system. As previously recalled, there are two main classes of ransomware, i.e. the \emph{locker} ones and the \emph{crypto} ones. In the first case, the goal of the ransomware is to lock the user's computer, using simple or sophisticated mechanisms, as to make it unable for the user to regain access to the computer. Then, they typically display a message across the screen to demand payment. Access is granted again only if the user pays the ransom. An early example of this class is \emph{WinLock}, which requested the ransom payment to be made through an SMS, whose cost was set to 10\$, to get the unblock code. On the other hand, crypto-ransomware aims to quietly search for and encrypt users' files, and then asks users to pay a ransom to get the decryption keys to access them again. Very often crypto-ransomware does not encrypt the whole hard-disk, but searches for specific extensions only, e.g. \texttt{.doc}, \texttt{.jpg}, \texttt{.pdf}, often of files containing text documents, presentations, images, which usually contain valuable and personal user data -- the ones that affect most the users if lost. Usually, symmetric encryption is chosen, for efficiency reasons, but asymmetric encryption as well as hybrid mechanisms, e.g. using a symmetric key to encrypt the files and using a public key to encrypt the symmetric keys \cite{0x3a}, have also been used. In both cases, ransomware makes extensive use of social-engineering tactics to pressure the victims into paying, for instance by masquerading themselves as law enforcement authorities, and claiming to issue fines for users for allegedly found criminal activities on the computer (e.g. illegal music files). Similarly, the ransomware threatens to publish the users' pictures and other personal data on the Internet unless a ransom is paid \cite{booming_ransomware}. According to a study by Symantec \cite{Symantec_tr_05_2016}, from 2005 to 2014 there were only 16 discovered families of Ransomware in the wild, a small number in comparison to the 27 discovered in 2015 alone and the 15 discovered only in Q1 of this year. Currently, the top ransomware families  are \emph{Cerber}, \emph{Locky}, and \emph{CryptoWall} \cite{statistics_ransomware}, while just few months ago they also included \emph{TorrentLocker}, \emph{CTB-Locker} and \emph{TeslaCrypt} (recently dismissed). This trend shows the continuously evolving nature of ransomware. A timeline of some notable ransomware is shown in Fig~\ref{tab:timeline}.

\begin{table*}[!hbt]
\centering
	\ra{1.1}
	\caption[]{Timeline of Representative Ransomware \label{tab:timeline}}
	%\resizebox{0.99\textwidth}{!}{
		\begin{tabular}{@{}lll@{}}
				\toprule
			\textbf{Name} & \textbf{Year} & \textbf{Notable Features} \\
			\midrule
PC Cyborg & 1989 & Spreads using floppy disks\\
GPCoder & 2005-2008 & Spreads via emails; encrypts a large set of files\\
Archiveus & 2006 & First Ransomware to use RSA encryption\\
WinLock & 2010 & \makecell[l]{Blocks PCs by displaying a ransom message}\\
Reveton & 2012 & \makecell[l]{Warning purportedly from a law enforcement agency}\\
DirtyDecrypt & Summ. 2013 & Encrypts eight different file formats\\
CryptLocker & Sept. 2013 & \makecell[l]{Fetches a public key from the C\&C}\\ % (ended May 2014)
CryptoWall & Nov. 2013 & Requires TOR browser to make payments\\
Android Defender & 2013 & First Android locker-ransomware\\
TorDroid & 2014 & First Android crypto-ransomware\\
Critroni & July 2014 & Similar to CryptoWall\\
TorrentLocker & Aug. 2014 & \makecell[l]{Stealthiness: indistinguishable from SSH connections}\\
%\textbf{CTB-Locker} & Dec. 2014 - Feb. 2015 & Uses Elliptic Curve Cryptography, TOR and Bitcoins Payment\\
CTB-Locker & Dec. 2014 & Uses Elliptic Curve Cryptography, TOR and Bitcoins\\
CryptoWall 3.0 & Jan. 2015 & Uses exclusively TOR for payment\\
TeslaCrypt & Feb. 2015 & Adds the option to pay with PayPal My Cash Cards\\
Hidden Tear & Aug. 2015 & Open source ransomware released for educational purposes\\
Chimera  & Nov. 2015 & Threatens to publish users' personal files\\
CryptoWall 4.0 & Nov. 2015 & Encrypts also filenames\\
Linux.Encoder.1 & Nov. 2015 & Encrypts Linux's home and website directories\\
DMA-Locker & Jan. 2016 &  Comes with a decrypting feature built-in\\
PadCrypt & Febr. 2016 & Live Chat Support\\
Locky Ransomware & Febr. 2016 & Installed using malicious macro in a Word document\\
CTB-Locker for WebSites & Febr. 2016 & Targets Wordpress \\
KeRanger & Mar. 2016 & First ransomware for Apple's Mac computers\\ 
Cerber & Mar. 2016 & Offered as RaaS (\& quote in Latin)\\
Samas & Mar. 2016 & Pentesting on JBOSS servers \\
Petya & Apr. 2016 & Overwrites MBT with its own loader and encrypts MFT\\
Rokku & Apr. 2016 & Use of QR code to facilitate payment\\
Jigsaw & Apr. 2016 & Press victims into paying ransom \\
CryptXXX  & May 2016 & Monitors mouse activities and evades sandboxed environment\\
Mischa & May 2016 & Installed when PETYA fails to gain administrative privileges\\
RAA & June 2016 & Entirely written in Javascript \\
Satana & June 2016 & Combines the features of PETYA and MISCHA\\
Stampado & July 2016 & Promoted through aggressive advertising campaigns on the Dark web \\
Fantom & Aug. 2016 & Uses a rogue Windows update screen\\
Cerber3 & Aug. 2016 &  Third iteration of the Cerber ransomware\\
\bottomrule
\end{tabular}
%}
\end{table*}

Technological advances have made ransomware more powerful. For example, virtual currency is now the \emph{de facto} method to pay ransoms for its untraceable properties \cite{bitcoins_main}. Usually, in these cases, a new, and unique, Bitcoin address is created for each user, so it can be used as a victim ID as well as to receive payments.  Furthermore, anonymous networks, such as Tor, are widely used to hide the location of the attacker's servers and to store the victim's private keys. In this context, CTB-Locker is probably the most advanced ransomware to-date \cite{CTBLOCKER}. Finally, note that to build a ransomware a cyber-criminal does not need to have sophisticated knowledge, as online services, such as \emph{ransomware-as-a-service} (RaaS), enable everyone to build their own version easily \cite{raas}.

\subsection{Infection Vectors}
To install ransomware on user's computer, several techniques are frequently used by cyber-criminals, such as:
\begin{itemize}
\item \textbf{Phishing or SPAM e-mails}: e-mails including a link \cite{sans_isc} (e.g., to Dropbox \cite{chimera}), or an attachment (using common words for the filename, such as 'invoice', 'internal' \cite{CTA_ransomware}) to a malware. Sometimes \texttt{.src} (Microsoft Windows Screensaver) files are disguised as \texttt{.pdf} with a fake icon, to lure victims into opening it; in any case, a malware sample is downloaded, and this malware downloads or drops the ransomware sample. Similar techniques apply to IRC, p2p networks, etc.
\item \textbf{Exploit kits}: rogue advertisements (ADs) are injected into an advertising network (\emph{malvertising}), usually by placing such ADs on reputable websites to enlarge the potential audience; the rogue AD redirects users to attacker's website, which usually exploits a vulnerability in the browser, using an exploit kit (e.g., Magnitude, Angler, Neutrino, and Nuclear \cite{exploit_kit}), to enable the drive-by-download of ransomware (or of malware downloading a ransomware in a second step). Note that almost 75\% of exploits in Angler are related to Adobe Flash, whereas 20\% are related to Internet Explorer \cite{talos_ransomware}.
\item \textbf{Downloader and Trojan Botnets}: downloaders coming from software-hosting websites, whose official goal is to allow users to download legitimate files, and as a hidden functionality download malware without the user noticing it.
\item \textbf{Social engineering tactics}: e.g., by deceiving users into installing a fake AV, by showing results of AV scans allegedly showing malware on the user's computer.
\item \textbf{Traffic Distribution Systems (TDS)} \cite{tds}: similar to malvertising, cyber-criminals buy redirected web traffic to the site hosting the exploit kit as to enable the drive-by-download of the malware.
\end{itemize}

For example, in CryptoWall Version 3, the two primary distribution channels have been phishing e-mails and exploit kits, where about two-thirds were phishing e-mails \cite{CTA_ransomware}. However, since April 2015, attackers began relying more heavily on exploit kits for distribution and propagation of CryptoWall, due to their flexibility and power \cite{CTA_ransomware}. The Angler Exploit Kit, the number one crime-kit to distribute CryptoWall, can actually inject its payload directly into the memory of infected machines and handily exploits a variety of vulnerabilities, especially in Flash.

\subsection{Recovering from Ransomware}\label{ref:recovering}
The recovery of a system compromised by locker-ransomware can be usually done, for example, by rebooting the system in safe-mode, and running an on-demand virus scanner, or through similar system-restore techniques. On the other hand, recovering the data from crypto-ransomware is usually a more challenging task. In fact, unless the user pays the ransom, and provided cyber-criminals are willing to give in exchange the key (which is usually the case, as they want to preserve their ``reputation''), in general it is very hard to recover the encrypted data (i.e., computationally infeasible). Usually, the private key is either stored on the server, or it is stored encrypted on the user storage (sometimes on the header of the encrypted files themselves) with public key mechanism. Obliviously, in case the encryption mechanism is not implemented correctly, e.g. cyber-criminals design and develop their own encryption algorithm, or the key length is too short, or there are flaws in the protocol \cite{foiled_ransomware}, it is possible to recover the data. However, note that usually this is not the case as ransomware, such as TeslaCrypt 2.0 exploits very sophisticated key generation mechanisms \cite{teslacrypt_20}. In some cases, the private keys is stored in clear (i.e., not encrypted with a public key) on the victim's disk, and they can be recovered. For example, in some GPcoder variants, the keys are stored in a registry key, and, hence, it is possible to recover the key. In other cases, files can be recovered by simply rebooting the machine before the ransomware terminates, as sometimes files are zeroed only after completing the encryption. We have also to consider those cases where the encryption keys is (by mistake) deleted after use, and not stored in the attacker's server \cite{ransomware_fail}. In these case, the only way users can recover their files is if they restore them from a backup. Ransomware may also sometime delete shadows copies containing old copies of files \cite{shadow_copy}. Note that some forensics techniques might be able to recover some files (or retrieve the key) or, similarly, sometimes cyber-criminals might disclose the private keys, e.g. after regretting their actions \cite{decryption_published}. Alternatively, the law enforcement might be able to access the database of encrypting keys or, in some other case, some AV vendors have made available an online repository of recovered decryption keys, e.g. for CoinVault and Bitcryptor\footnote{\url{https://noransom.kaspersky.com/}}, or an online repostory of several decrypters for free\footnote{\url{https://decrypter.emsisoft.com/}}.

%---analysis----%
\section{\ourname{}: Classification \\ of Ransomware}\label{sec:analysis}
In this section we discuss our approach, and motivate the selection and use of features to detect ransomware.
\subsection{System Design} 
\begin{figure*}[!hbt]
\centering
\includegraphics[width=0.5\textwidth]{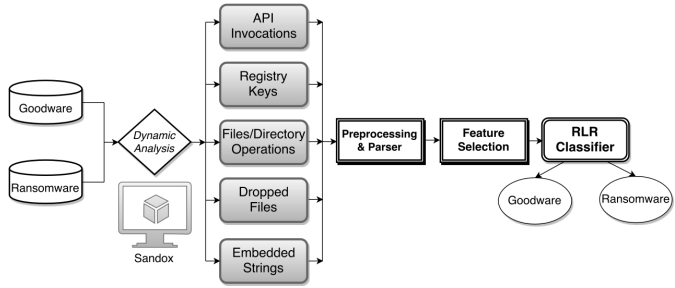}
\caption{Elderan Sandboxed Training and Analysis\label{fig:elderan}}
\end{figure*}

\begin{figure*}[!hbt]
\centering
\includegraphics[width=0.5\textwidth]{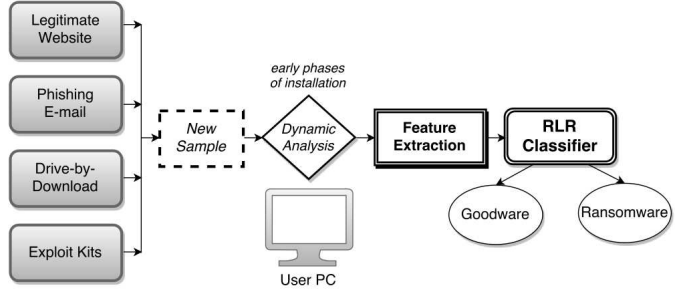}
\caption{Elderan Live Detection\label{fig:elderan2}}
\end{figure*}

\ourname{} is based on the observation that ransomware samples typically perform some actions that are unique, or significant, with respect to those performed by goodware ones. Therefore, \ourname{} is focused on deriving the most important features of ransomware in its early phases. Figures~\ref{fig:elderan} and ~\ref{fig:elderan2} show, respectively, the design of our proposed system in the analysis and detection phases. In detail, \ourname{} firstly dynamically analyses in a sandboxed environment (see Fig.~\ref{fig:elderan}) the traces of samples coming from two datasets of ransomware and goodware. From these datasets, \ourname{} retrieves, and analyses, the following classes of features: (i) Windows API calls (i.e., the traces of invocations of native functions and Windows API calls), (ii) Registry Key Operations (in particular, the read, open, write, and delete operations), (iii) File System Operations (in particular, the read, open, write, and delete operations), (iv) the set of file operations performed per File Extension, (v) Directory Operations (the set of operations performed on directories, in particular the enumeration and creation), (vi) Dropped Files (i.e., the set of files that are dropped by an application during installation), and (vii) Strings (the strings embedded in the binary). All the features, but the last ones (Strings) are collected while dynamically analysing the ransomware. After this monitoring phase, a feature selection algorithm is applied to select the most relevant ones. Finally, the matrices containing these features are used in a Regularized Logistic Regression classifier, which returns ``ransomware'' or ``goodware'' for, respectively, a ransomware or a good application. The classifier is also run online on user PCs to classify new samples, coming from genuine (compromised) website or from various infection vectors (see Fig~\ref{fig:elderan2}). Note that while the training set is analysed offline, and its classification is done \emph{una tantum} on a sandboxed environment and requires few minutes only, new applications are classified at run-time through an online classifier, which is very fast and can be run on user PCs.

\subsection{Feature Selection and Classification}\label{sec:classifier}
The machine learning component of \ourname{} consists of two phases: feature selection and classification. For the first one, we have used the Mutual Information criterion \cite{cover}, which allows us to select the most discriminating features for ransomware and goodware from the features gathered in the sandbox. Then, for the classification (decision) problem, \ourname{} uses a Regularized Logistic Regression classifier.

\paragraph{Feature Selection}\label{sec:feature_selection}
Feature selection is a procedure that enables reducing the number of features to build simpler machine learning algorithms, by shortening the training (and prediction) time, and, in many cases, to prevent overfitting. Even if these techniques are not always used in machine learning approaches to detect malware \cite{kolter2006learning,Rieck:2011:AAM:2011216.2011217}, in our opinion they are a key element to make the algorithm more efficient and achieve a better performance, as we will show in Sect. \ref{sec:experiments}. In \ourname{}, for each feature, we consider only the presence or absence of that particular feature.
Hence, for these binary features, the Mutual Information criterion is a good method to select the most relevant ones, since it allows us to quantify how much discrimination each feature adds to the classifier. Considering that $X$ and $Y$ are two discrete random variables with a joint probability mass function $p(x,y)$ and marginal probability mass functions $p(x)$ and $p(y)$, the Mutual Information $\text{MI}(X,Y)$ is defined as the relative entropy between the joint distribution and the product of the marginal distributions \cite{cover}:
\begin{equation}
\text{MI}(X,Y) = \sum_{x \in {\cal X}}  \sum_{y \in {\cal Y}} p(x,y) \log \frac{p(x,y)}{p(x) \ p(y)}
\label{eqMI}
\end{equation} 

\paragraph{Regularized Logistic Regression}\label{sec:rlr}
Once we have applied the feature selection, the final step is to build a classifier to detect ransomware. Since in this case we have a high number of features, linear classifiers are usually a good choice, as shown in \cite{kolter2006learning,shafiq2009pe} where different linear classifiers were proposed to detect malware. In our case, we have resorted to the \textit{Logistic Regression} classifier. This method aims to model the log-posterior probability of the different classes given the data via linear functions depending on the features \cite{hastie}. Then, the posterior probability of a sample being classified as ransomware ($R = 1$) given its feature vector ${\bf x}$ can be written as:
\begin{equation}
p(R = 1 | {\bf x}, w, b) = {\text{sgm}} \left( {\bf w}^\top {\bf x} + b \right)
\label{eqPosterior}
\end{equation} where ${\bf w}$ is the vector of weights, $b$ is the bias term, and the sigmoid function, $\text{sgm}(t)$ is given by:
\begin{equation}
{\text{sgm}}(t) = \frac{1}{1 + \exp(-t)}
\label{eqSigmoid}
\end{equation}

To fit the values of ${\bf w}$ and $b$, we minimize a cost function between the estimated posteriors and the true labels on a set of training points. Although there are different cost functions that can be applied, one of the most common is the \textit{cross-entropy}, which also uses concepts from Information Theory. We selected this cost function since it is more sound when aiming to model the log-posterior probability of the classes although, in practice, we can expect a similar performance with other cost functions like the squared error. Therefore, given a set of $N$ training samples with $D$ features $\{ {\bf x}_i \}_{i=1}^N$, where ${\bf x}_i \in {\cal R}^D$, and the corresponding labels $\{ y_i \}_{i=1}^N$, where $y_i \in \{0,1\}$, the cross-entropy cost function can be written as:
\begin{equation}
\begin{split}
{\cal C} ({\bf w},b) = \frac{1}{N} \sum_{i = 1}^N y_i \log p(R = 1 | {\bf x}_i, w, b) + \\ (1 - y_i) \log \left( 1 - p(R = 1 | {\bf x}_i, w, b) \right)
\end{split}
\label{eqCE}
\end{equation} The cost function is convex on the weights and the bias, which means that it has a unique global minimum. Hence, to search the values of ${\bf w}$ and $b$ that minimize (\ref{eqCE}) we can use gradient descent methods. In our case, given the size of the dataset, we have opted for a standard gradient descent approach in batch mode, i.e. computing the gradient for the whole dataset at each iteration. For bigger datasets, or to perform \textit{online} training, stochastic gradient descent or mini-batch gradient descent may be more appropriate.

One of the main limitations of the Logistic Regression is that the maximum likelihood estimation of the posterior $p(R = 1 | {\bf x}, w, b)$ is prone to overfitting. However, several techniques exist in the literature to prevent this problem. Although, we can resort to Bayesian models, placing a prior on ${\bf w}$ and $b$ \cite{bishop}, a simpler and common alternative that achieves similar performance to Bayesian models is regularization \cite{hastie}. Hence, we can prevent overfitting in the Logistic Regression model by adding a penalty term to the cost function (\ref{eqCE}). In our case, we have used an $L_2$ regularization term, so that the cost function for the Regularized Logistic Regression becomes:
\begin{equation}
{\cal C}' ({\bf w},b) = {\cal C} ({\bf w},b) + \frac{\lambda}{2} \sum_{i=1}^D w_i^2
\label{eqRCE}
\end{equation} where $\lambda$ is the regularization parameter, which is typically set by cross-validation. Note that in \ourname{} we have two mechanisms to prevent overfitting: the feature selection with the Mutual Information criterion and the regularization of the Logistic Regression. 

Although more powerful classifiers, like the SVM, could have been used, in many practical situations the Regularized Logistic Regression is competitive with SVM and is easier to train and to adapt when new training samples are added to the training set. Hence, the parameters can be adapted online as new samples arrive without the need of training the classifier from scratch. On the other hand, although in some practical situations the simplistic approach of the Naive Bayes classifier achieves a good performance, it relies on the assumption of independence between features, which we do not consider to be valid for the detection of ransomware, where we expect to have strong dependencies between some of the features, e.g. between API and File Operations.
In Sect.~\ref{sec:experiments} we will show the experimental comparison of the performance the Regularized Logistic Regression classifier compared to the SVM and Naive Bayes.

%---dataset----%
\section{Analysis of the Features} \label{sec:dataset}
The ransomware samples of our datasets were downloaded from VirusShare\footnote{\url{http://virusshare.com/}}, a website that maintains a continuously-updated database of malware for several OSes. Table~\ref{tabRansomware} reports the full list of ransomware families used for our experiments. To analyse the samples we used the Cuckoo Sandbox\footnote{\url{http://www.cuckoosandbox.org/}}, a well-know tool to automate malware analysis. The construction of the dataset, and the use of the sandbox, have been done following the best practices suggested in \cite{6234405}, and is described in the following. %before it was an appendix
Later, we describe the features we extracted and processed.

\subsection{Dataset and Sandbox} \label{ref:app_cuckoo}
The dataset of ransomware and goodware was retrieved and analysed at the end of February 2016, and it consists of 582 working samples of ransomware belonging to 11 different classes and 942 of good applications\footnote{The initial dataset was larger: we considered a ransomware or a goodware sample to be effective only if the set of dynamic features was not empty.}.
We strived to collect both old ransomware and new ransomware samples, as well as recent and well-known good applications (see in the following). The collected ransomware samples are representative of the most popular versions and variants currently encountered in the wild (most of them belong to crypto-ransomware type). We manually clustered each ransomware into a well-established family name, since there are several discrepancies among AV vendors' naming strategy and it is therefore not easy to extract a common name for each ransomware family. For each family, we downloaded all the samples found on VirusShare but, for the largest classes (having more that 100 samples), we limited the number of samples by randomly sampling from VirusShare database. We also downloaded and analysed a dataset of 942 benign applications (the \emph{goodware} set): these were selected from the top list of the most popular software coming from a software aggregator website\footnote{\url{http://software.informer.com/software/}}. To ensure that the downloaded goodware samples did not contain suspicious components inside their payload, we only downloaded applications hosted from the most trustworthy sources only, as indicated by the internal AV checker and rating system. Finally, we also double-checked with the results of VirusTotal. This dataset of goodware applications includes generic utilities for Windows (e.g., zipper, password managers, etc.), drivers, browsers (the most popular ones), file utilities (DropBox, file search, etc.) multimedia tools (music, video, etc.), developers tools (Eclipse, notepad++, etc.), games, network utilities, paint tools, databases, emulator and virtual machines monitors, office tools, etc. We tried to built a realistic dataset that encompasses a large variety of common applications installed on user PCs. We manually checked that both classes did not contain samples of the other class, i.e. ransomware labelled as goodware and viceversa. 

The system used during execution was a version of Windows XP 32 bit SP2 without additional software installed, except for Python (which was requested by Cuckoo for its agent). We used this version of Windows as its weaker security protections enables EldeRAN to observe more ransomware behaviour \cite{184519}. When analysing the malware execution, no other user application was running in the analysed machine, and the system was not actively used by people. The virtual machine (VM) running in the sandboxed environment with Windows XP had network connectivity. We also inserted several files, under different directories, to mimic personal user files, such as images (e.g., jpg, png), Office documents (e.g., doc, ppt), documents (e.g., pdf, txt), and others (e.g., zip). Each VM was reverted to the original clean/safe state before each new test. We analysed the samples of ransomware and goodware in the Sandbox for 30 seconds.  Although we allowed network connectivity to the VM, and collected PCAP traces, we focus here on the analysis of host-based features only.

\subsection{Features}

To analyse all the dinamically-generated traces efficiently, we built a \emph{feature parser} (see ``preprocessing \& parser'' component in Fig.~\ref{fig:elderan}) that retrieves all these reports (in JSON format), and creates a set of matrices that can be analysed by the EldeRAN feature selection and classifier. Before creating the matrices, the EldeRAN feature parser implements some preprocessing and basic feature selection. In detail, in case of API invocations, the feature parser creates a matrix with the \emph{API Stats} feature, which takes into account the absence or the presence of a specific call for this feature.
Concerning the registry keys (\emph{Reg Keys}), for each of the four analysed operations, and for each sample, the feature parser creates a matrix by considering whether the operation was performed on each specific key accessed by the operation.
Concerning the \emph{File Operations}, the feature parser creates similar matrices to the ones related to the registry keys by considering as features the full paths of the filename along with the absence or presence of the specific operation.
Analogously to File Operations, the features parser creates similar matrices by considering the file operations performed per \emph{File Extension}, and by considering the \emph{Directory Operations} (enumerate and creation).
For the \emph{Dropped Files} feature, the feature parser considers as feature the type of the dropped file, i.e. a new feature is added for each new type of dropped file (such as PE32, DOS batch, JPEG, etc.), coupled with the absence of presence of each feature for all the samples of ransomware and goodware.
Finally, for the \emph{Strings} features, since the set of embedded strings is fairly large, we limited ourselves to consider as features each imported library and function, as well as strings related to crypto-functions, and string related to file extensions, by creating a single matrix considering the absence or presence of each of these features. After analysing all the samples of ransomware and goodware, the total number of features was $30,967$.
We then further performed experiments to assess which class of features is better suited to classify ransomware when isolated. By considering only the top 400/100 features, according to the MI criterion, the percentage of features belonging to each set is shown in Tab.~\ref{tab:relevantFeatures}. 

\begin{table}[!hbt]
\centering
\ra{1.1}
\caption[]{Percentage of the Most Relevant Features for Each Class \label{tab:relevantFeatures}}
\begin{tabular}{@{}lcc@{}}
\toprule
Features & Top 400 & Top 100\\
\midrule
Registry Keys Operations & 48.25\% & 49\% \\
API Stats & 24.00\% & 27\% \\
Strings & 8.25\% & 5\% \\
File Extensions & 8.00\% & 9\% \\
Files Operations & 5.25\% & 6\%\\
Directory Operations & 4.00\% & 2\% \\
Dropped Files Extensions & 2.25\% & 2\% \\
\bottomrule
\end{tabular}
\end{table}

We observe that the Registry Keys and API Stats are the two most relevant sets, but also the other sets are useful. Among all these features, several are characteristic of ransomware behaviour, and their presence, together with other features more generally characteristic of malware behaviour, leads to a very good detection rate (see Sect.~\ref{sec:experiments}). In particular, high activities on files in personal folders and tmp directories. Ransomware also often implement registry keys updates to enable persistence across reboot, or access to keys to retrieve the list of mounted devices (to search for more files), or to acquire the list of recently opened files, possibly because believed to contain important information. Further, ransomware often include links to DLL related to the usage of Visual Basic, shell extensions, and access to certificates (related to key creation and management), or strings showing patterns of extension to be searched for. Characteristic APIs are those allowing a ransomware to terminate, or write to, other processes (for example, to inject into \texttt{explorer.exe} and \texttt{svchost.exe} processes). Dropped files include html and rtf, used to show the ransom note.  The usage of evasion techniques seems to be recurrent, e.g. trying to search for the Python agent.  Finally, important file extensions include manifest files, pad, and bat.

%---experiments----%
\section{Experiments} \label{sec:experiments}
We present here the experimental evaluation to assess the accuracy of \ourname{} to detect ransomware, by measuring the performance and the detection rate of the classifier compared to other machine learning techniques and AV vendors. We used a Naive Bayes (NB) classifier \cite{hastie} and the linear SVM \cite{boser,vapnik} as comparison benchmarks to assess the performance of the Regularized Logistic Regression classifier used by \ourname{}. We used the Matlab (version 2015b) implementation for the Naive Bayes and the SVM, whereas for EldeRAN we developed our own Matlab implementation. We applied the MI criterion to select the most relevant features for the three methods. In the experimental evaluation we also included the information provided by VirusTotal with the labels from a pool of AV vendors indicating if a sample is malware or not. To this end, we have aggregated the labels from the pool of AV providers using a majority voting strategy. Then, we decided that a sample is ransomware if more than a half of the AV vendors providing a label for that sample says that it is malware. For the experimental comparison, we also considered individually the 5 AV vendors with the highest detection rate measured on the whole dataset. In the cases where an AV vendor did not provide a label for a given sample, we did not consider that sample when calculating the FPs and detection rates for that AV vendor. Note that this induces an experimental advantage for the AV vendors. Finally, we have also analysed the ability of \ourname{} to detect new families of ransomware, i.e. families that are not included in the features selection, and the training process of the classifier. 

\subsection{Comparison with other Classifiers}
For the first experiment we evaluated the performance of the 3 machine learning classifiers (\ourname{}, SVM, and NB) with the number of features selected with the MI criterion. We explored using between $50$ and $1,500$ features to see which combination performs better. For each value explored, we created $100$ random splits of the whole dataset with $80\%$ of the samples for training and $20\%$ as test samples. To select the regularization parameter $\lambda$ for \ourname{} and the cost parameter $C$ for the SVM, we performed a 5-fold cross validation using the whole dataset, exploring values for $\lambda$ and $C$ in the range $[2 \cdot 10^{-5}, 2 \cdot 10^{-1}]$ with values of the form $2 \cdot 10^p$, with $-5 \leq p \leq -1$. For \ourname{} we also explored the value $\lambda = 0$, which corresponds to the standard Logistic Regression, i.e. without regularization. To train the Regularized Logistic Regression we used a standard gradient descent implementation in batch mode, with a learning rate of $0.8$, and with $4,000$ as the max number of training iterations.

\begin{figure}[!hbt]
	\centering
	\psfrag{AUC}{{\scriptsize AUC}}
	\psfrag{Features}[-0.3cm]{{\scriptsize \# of features}}
	\psfrag{EldeRAN}{{\scriptsize EldeRAN}}
	\psfrag{SVM}{{\scriptsize SVM}}
	\psfrag{Naive Bayes}{{\scriptsize Naive Bayes}} 
	\includegraphics[width=8cm,height=5cm]{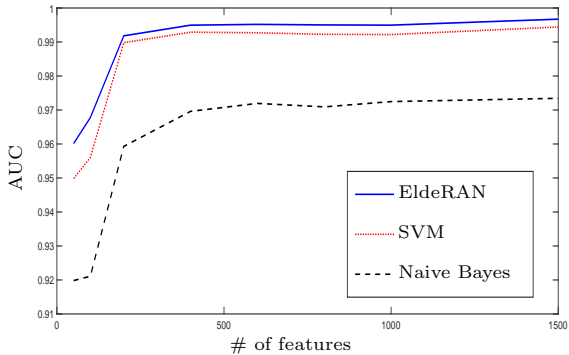}
	\caption{Evolution of the average AUC with the number of features over 100 train/test splits for \ourname{}, the SVM, and NB}
	\label{figAUC}	
\end{figure}

In Fig. \ref{figAUC} we show the performance of \ourname{}, the SVM, and NB in terms of the AUC with the number of features. We observe that both \ourname{} and the SVM outperforms the NB classifier. This result suggests the importance of considering the relation across the different features. Note that NB considers independence between features, whereas the SVM and the Logistic Regression take into account possible dependencies across features. We observe that \ourname{} slightly outperforms the SVM in all cases. Although the difference is not significant, the Regularized Logistic Regression classifier used by \ourname{} is simpler than the SVM and can be easily adapted to be retrained online, i.e. as we increase our training dataset with new samples, without training the classifier from scratch. This property is desirable from a practical perspective since we can expect  hundreds of new ransomware and goodware samples per day.  By looking at Fig.~\ref{figAUC}, it is also interesting to observe that the maximum of the performance for the three algorithms is achieved for $400$ features. Thus, adding more features to the classifiers does not improve the accuracy. This result shows the importance of feature selection to reduce the complexity of the machine learning algorithms without affecting the performance (note that the whole number of features in our dataset is $30,967$). Moreover, this also suggests that MI is an effective means of identifying the most useful features. Therefore, for the remaining experiments we used the $400$ features with the highest MI for \ourname{}, the SVM, and NB.

\begin{figure}
	\centering
	\psfrag{PFA}[-0.2cm]{{\scriptsize False Alarm}}
	\psfrag{Detection}[-0.2cm]{{\scriptsize Detection}}
	\psfrag{EldeRAN}{{\scriptsize EldeRAN}}
	\psfrag{SVM}{{\scriptsize SVM}}
	\psfrag{Naive Bayes}{{\scriptsize Naive Bayes}} 
	\psfrag{VirusTotal}{{\scriptsize VirusTotal}} 
	\includegraphics[width=8cm,height=5cm]{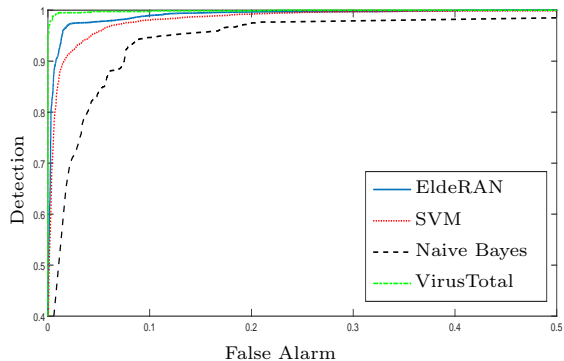}
	\caption{Average ROC for the test samples over 100 random splits for \ourname{}, the SVM, NB, and VirusTotal}
	\label{figROC}	
\end{figure}

\subsection{Comparison with VirusTotal}
For the second experiment we compared the performance of \ourname{}, the SVM, and NB with VirusTotal.
For \ourname{}, the SVM, and NB we selected the top $400$ features.
We have computed the test performance for all the methods averaging the results over $100$ independent train/test splits with $80\%$ of samples for training. In Fig. \ref{figROC} we show the average ROC for the test set. We can observe that VirusTotal outperforms the rest of the machine learning algorithms, though \ourname{} comes close to its performance. On the other hand, \ourname{} outperforms both the SVM and NB, supporting the results of Fig. \ref{figAUC}. In Tab. \ref{tabResults1} we show a more detailed comparison of the four methods in terms of the AUC and test error, false positive, and detection rates. We observe that in terms of the AUC, VirusTotal slightly outperforms SVM and \ourname{}, which supports the results in Fig. \ref{figROC}. However, in terms of the test error rate VirusTotal (with majority voting) is worse than \ourname{} and the SVM. In this sense, whereas \ourname{}'s average error rate is $2.4\%$, that of VirusTotal is $5.6\%$. We also observe a better performance of \ourname{} with respect to the SVM (with a $4.2\%$ error rate) and NB ($8.0\%$ of error). From the results in Tab. \ref{tabResults1} it is interesting to note that in the case of VirusTotal all the errors are false negatives, since the false positive rate is zero. This implies that the detection rate of VirusTotal is sensibly lower compared to the 3 machine learning techniques. \ourname{} outperforms the rest of the methods and achieves a remarkable $96.3\%$ detection rate. Although the false positive rate is slightly higher than VirusTotal, we consider that a $1.6\%$ of false positives are tolerable for this kind of analysis, considering the fact that \ourname{} is not based on signatures, and is also able to detect new ransomware families. The good average ROC for VirusTotal in Fig. \ref{figROC} contrasts the poor results in terms of the detection rate.

\begin{table*}[!hbt]
\centering
\ra{1.1}
\caption[]{Average test results plus/minus one standard deviation for \ourname{} (ELD), the SVM, NB, and VirusTotal (VT) over $100$ independent train/test splits, including the AUC, test error, false positive (FP), and detection rates \label{tabResults1}}
\setlength{\tabcolsep}{0.2cm}
\begin{tabular}{@{}lcccc@{}}
\toprule
Method & AUC & Test Error & FP Rate & Det. Rate \\
\midrule
ELD & $0.9949 \pm 0.0025$ & $0.0238 \pm 0.0090$ & $0.0161 \pm 0.0088$ & $0.9634\pm 0.0215$  \\
SVM & $0.9929 \pm 0.0033$ & $0.0421 \pm 0.0108$ & $0.0199 \pm 0.0107$ & $0.9219 \pm 0.0244$ \\
NB & $0.9696 \pm 0.0081$ & $0.0800 \pm 0.0118$ & $0.0958 \pm 0.0162$ & $0.9453 \pm 0.0212$ \\
VT & $0.9993 \pm 0.0008$ & $0.0561 \pm 0.0120$ & $0.0000 \pm 0.0000$ & $0.8530 \pm 0.0301$ \\
\bottomrule
\end{tabular}
\end{table*}

\begin{table}[!hbt]
\centering
\ra{1.1}
\caption[]{Average test results plus/minus one standard deviation (over $100$ train/test splits) for \ourname{} (ELD) compared with the Top-5 AV vendors, including the test error, false positive (FP), and detection rates \label{tabResults2}}
\setlength{\tabcolsep}{0.2cm}
\begin{tabular}{@{}lccc@{}}
\toprule
 & Test Error & FP Rate & Det. Rate \\
\midrule
ELD & $0.0238 \pm 0.0090$ & $0.0161 \pm 0.0088$ & $0.9634 \pm 0.0215$  \\
AV1 & $0.0159 \pm 0.0060$ & $0.0066 \pm 0.0048$ & $0.9689 \pm 0.0142$ \\
AV2 & $0.0274 \pm 0.0082$ & $0.0196 \pm 0.0080$ & $0.9596 \pm 0.0162$ \\
AV3 & $0.0165 \pm 0.0064$ & $0.0000 \pm 0.0000$ & $0.9567 \pm 0.0167$ \\
AV4 & $0.0200 \pm 0.0066$ & $0.0000 \pm 0.0000$ & $0.9469 \pm 0.0173$ \\
AV5 & $0.0205 \pm 0.0079$ & $0.0000 \pm 0.0000$ & $0.9426 \pm 0.0266$ \\
\bottomrule
\end{tabular}
\end{table}

\begin{figure}
	\centering
	\psfrag{EL}[-0.05cm]{{\scriptsize ELD}}
	\psfrag{A1}[-0.05cm]{{\scriptsize AV1}}
	\psfrag{A2}[-0.05cm]{{\scriptsize AV2}}
	\psfrag{A3}[-0.05cm]{{\scriptsize AV3}}
	\psfrag{A4}[-0.05cm]{{\scriptsize AV4}} 
	\psfrag{A5}[-0.05cm]{{\scriptsize AV5}}
	\psfrag{Detection Rate}[-0.2cm]{{\scriptsize Detection Rate}} 
	\includegraphics[width=8cm,height=5cm]{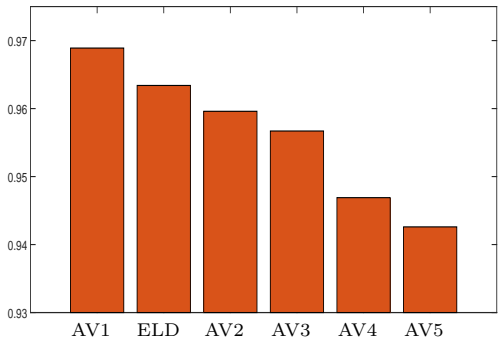}
	\caption{Average detection rate (over $100$ train/test splits) of \ourname{} compared with the Top-5 AV vendors}
	\label{figDetection}	
\end{figure}

Finally, given the limitations on the detection rate for the majority voting scheme with VirusTotal, we compared \ourname{} with the 5 AVs with highest detection rate over the whole dataset. Then, we averaged the results of these 5 AVs over the same $100$ train/test random splits used in the previous experiment. The results are shown in Tab.~\ref{tabResults2}. We can observe that \ourname{} outperforms 4 out the top-5 AVs in terms of the detection rate. Despite \ourname{}'s false positive rate is higher than 4 of the AVs, the test error rate is competitive to all of them. In Fig.~\ref{figDetection} we show in more detail the average detection rate of the top-5 AVs compared to \ourname{}, and we can see that \ourname{} places in the second position with a slightly lower detection rate compared to the best AV. As mentioned before, it is important to note that the experimental comparison is not fair, since machine learning methods and AV software work in a different way. In particular, the measured test performance corresponds to the performance on samples that the machine learning classifier has not seen yet, whereas possibly, in most cases, the AV vendors have the corresponding signature analysis for those test samples, i.e. they have already seen them. Moreover, when an AV does not provide an answer to a specific piece of software, we are not considering that as an error\footnote{Both when a vendor does not provide an answer for a sample on VirusTotal, and when the analysis is absent on VirusTotal. Note that we used the ``cached'' mode: hence, most of the goodware samples were not present on VirusTotal.}.

\subsection{Detection of New Ransomware Families}
In the final experiment we want to test the ability of \ourname{} to detect new families of ransomware. The importance of this analysis lies in the fact that often new families of ransomware appear, which share many characteristics, and goals, of previous classes \cite{dimva_paper}. Therefore, particular in the first phases, the average detection of these signatures by AV vendors is very low, exemplified by the CryptoWall 3 variants \cite{crypto_3}. We manually clustered our dataset into $11$ classes using their generally known family name -- as the naming conventions of the AV vendors are not always consistent or compatible amongst them.
We analysed the detection rate of \ourname{} when one specific family of ransomware is not considered in the training set, so that family is effectively a new family for the classifier. Then, for each of the $11$ ransomware families appearing in the dataset, we selected the features and trained the classifier considering all the goodware and ransomware samples in the dataset, except those corresponding to the ransomware family under analysis. We considered two cases by selecting the top $100$ and $400$ features according to the MI criterion. For the classifier, we set regularization parameter $\lambda = 2 \cdot 10^{-3}$. When testing the detection rate for each family, since the samples in the training set are the same, we do not need to average the results over several repetitions. Note that, for a given training set, the Logistic Regression classifier converges always to a unique minimum of the cost function\footnote{In practice there can be very small variations due to the gradient descent algorithm, but they are negligible.}.

\begin{table}[!hbt]
\centering
\ra{1.1}
\caption[]{Detection rate across different ransomware families when \ourname{} is trained using only the data from the rest of the ransomware families using the most relevant $100$ and $400$ features according to the Mutual Information criterion \label{tabRansomware}}
\setlength{\tabcolsep}{0.2cm}
\begin{tabular}{lccc}
\toprule
Family & \# samples & \makecell{Det. Rate \\($400$ ft.)} & \makecell{Det. Rate \\ ($100$ ft.)} \\
\midrule
Citroni & 50 & $0.9200$ & $0.9800$ \\
CryptLocker & 107 & $0.9065$ & $0.9626$ \\
CryptoWall & 46 & $0.7391$ & $0.9130$ \\
Kollah & 25 & $0.7600$ & $0.9600$ \\
Kovter & 64 & $0.8906$ & $0.8906$ \\
Locker & 97 & $0.8557$ & $0.9175$ \\
Matsnu & 59 & $0.9153$ & $0.9831$ \\
Pgpcoder & 4 &$1.0000$  & $0.7500$ \\
Reveton & 90 & $0.9111$ & $0.8889$ \\
TeslaCrypt & 6 & $0.8333$ & $1.0000$ \\
Trojan-Ransom & 34 & $0.7647$ & $0.9412$ \\
\hline
Weighted Avg. & 582 & $0.8711$ & $\mathbf{0.9330}$ \\
\bottomrule
\end{tabular}
\end{table}

The results of the experiment are shown in Tab. \ref{tabRansomware}. We can observe that, by using only $100$ features, the detection rate is above $90\%$ for $8$ of the ransomware families, and above $80\%$ for $10$ out of the $11$ families. The results are however worse when considering $400$ features with a detection rate higher than $90\%$ for $5$ families, and a detection rate above $80\%$ for $8$ families. It is interesting to observe that the average detection rate (calculated as the weighted average over the detection rate for each family) is higher ($93.3\%$) when using only the top $100$ features than in the case of using $400$ features ($87.1\%$). This contrasts the results shown in Fig. \ref{figAUC}, where for $100$ features we get a lower AUC and then, a lower detection rate is expected (at least if we consider the same false positive rate in the corresponding ROCs). The reason of this comes from the difference between the samples in the training and the test sets: whereas in the experiment shown in Fig. \ref{figAUC} the samples for both the training and the test sets come from the same distribution (since the samples from the different families are mixed in both sets), in the experiment shown in Tab. \ref{tabRansomware} the distribution of the ransomware samples in the test set can be different to the ones in the training set. This induces certain ``overfitting" to the families that are known to the classifier and limits the detection for new families of ransomware. However, in this particular case, it seems that all the ransomware families share a common set of features that are captured by the top $100$ most relevant features according to the MI criterion. This reduces the effect of the ``overfitting" due to the different distribution of the training and test sets. It is important to note that this ``overfitting" is not due to the Regularized Logistic Regression classifier and the feature selection algorithm \textit{per se}, but to the differences between the distributions of the training and test sets.

%---discussion----%
\section{Discussion} \label{sec:discussion}
In this section we discuss the results of \ourname{}, the benefits and limitations of dynamic analysis and suggestions for improvements.
\paragraph*{Benefits of Dynamic Analysis}
The  experimental results show that it is possible to detect a ransomware from its early phases of installation, and also when a new family appears. In fact, \ourname{} achieved a detection rate of 96.3\%. Note that we can expect to see more new ransomware families, especially with the RaaS availability.
In our experiments we have shown that, even if the initial set of features was rather large (around $31,000$), in the end only 400 have been shown to be sufficient. One important result that we were able to provide is that Mutual Information is very effective in deriving the most important features. In particular, the registry key and API calls are the two classes with most relevant features. Finally, \ourname{} is competitive with the top AV vendors in the detection rate. In this respect, note that security researchers have discovered ransomware variants sharing the same packing and delivery technique (essentially, they come from the same family) but included random variables in their code, for each target, that help them evading AV static signature-based detection \cite{kofer}, especially in the first phases of a new ransomware campaign, when as few as 2 AV vendors out of 66 may be able to detect a new variant \cite{low_detection}. This  example further favours our assumption that a machine learning approach is particularly suitable for detecting ransomware. Despite these very good results, \ourname{} would not be suitable as a replacement for AV software, and is not intended as such. However, these results indicate that \ourname{} would be an effective and entirely automated tool to analyse new software and enhance the detection capabilities of AV software. In particular, it shows that \ourname{} can identify ransomware with a high degree of accuracy, especially when new families of ransomware are spread on the wild, and can thus identify candidates for subsequent signature extraction.

Note that AV companies are relying on automated dynamic analysis tools to detect new variants of ransomware (and other malware): they apply heuristics combined with behaviour analysis to deduce whether the executable is benign or malware. \ourname{} follows a similar approach and, in addition, it implements machine learning techniques to improve the process of classification. There are two main reasons for employing automated dynamic analysis techniques for ransomware: first, in initial stage of triage, the main goal is the containment, i.e. the detection of the ransomware before infection; to this end, we run the infected file for a very limited time to select the relevant features; secondly, ransomware use very sophisticated packing (such as Themida and VMProtect) and obfuscated techniques that render most of static analysis tools useless. Therefore, dynamic analysis is indispensable to dissect ransomware and understand their main features and functionalities. Note that some families of ransomware do not drop their payload before checking certain preliminaries (environment, C\&C ), where in the absence of these the malware won't perform the encryption. Hence, early detection before releasing the payload in crucial for containment.

\paragraph*{Limitations of Dynamic Analysis}
We are aware that currently some limitations of \ourname{}.
A first limitation concerns the analysis, and detection, of those samples of ransomware that are silent for some time, or that wait for the user to do something. In this case, \ourname{} does not properly extract their features. To mitigate this ransomware's behaviour, a solution is to inject some real, or script-generated, user actions into the sandboxed environment. Analogously, for ransomware that exploits evasion, or anti-sandbox, techniques, e.g. any malware that looks for signs of emulation or virtualization, and that does not perform any action in case.  Another limitation of our approach is that, in the current settings, no other applications were running in the analysed VM, except the ones coming packed with a fresh installation of Windows. This might not be a limitation \emph{per se} but, as in the previous cases, the ransomware might do some checks as to evade being analysed. Finally, note that the dataset consists of 582 working samples of ransomware belonging to 11 different classes and 942 samples of good applications: however, the initial dataset was larger (1,450 ransomware samples and 1,131 goodware samples). We analysed a ransomware (or goodware) only if the set of API calls was not empty, which means the ransomware could be run properly. 

\paragraph*{Improvements}
In additions to the previously suggested improvements, we also point out that any dynamic solution should to take into account these issues: (i) reduce the time of feature analysis to be as short as possible; (ii) implement more sophisticated detection techniques, e.g. based on (known) patterns of system calls used to encrypt files; (iii) be able to distinguish between actions performed by a legitimated encryption software and a crypto-ransomware. We believe that, in any case, the dynamic analysis is needed to improve the detection of ransomware: in fact, current packers render the static analysis of ransomware very hard. Note that the goal of this work is twofold: firstly, to discover whether it is possible with dynamic analysis to derive some key-characteristics features of ransomware across different families. Secondly, whether it is possible, by using these features, to classify ransomware using machine learning techniques by achieving results comparable to those of top AV vendors. Note that the classifier, once trained, can be run on user PCs: as we already pointed out, the Regularized Logistic Regression classifier used by \ourname{} can be easily adapted to be retrained online when new samples are available, without the need to re-train it from scratch.

%---related----%
\section{Related Work} \label{sec:related}
Currently, little related work specifically targets ransomware. Hence, here we describe both related works on ransomware and more general malware-oriented papers. 
The first description of the development of a crypto-virus prototype was in 1996 \cite{502676}, which described the use of asymmetric encryption. \cite{virii} report the first analysis of three ransomware families, and showed they did not fulfil the basic crypto requirements (e.g., sufficiently long encryption keys) for mass extortion. \cite{young2006cryptoviral} present the experimental results of applying cryptography to carry out extortion-based attacks, by implementing the payload through Microsoft Cryptographic API. More recently, a Symantec report \cite{evolution_symantec} shows how ransomware has evolved considerably in the last years, both in terms of features and spreading capabilities. The authors also pinpoint future directions of ransomware evolution, such as targeting the wearable market (the so called ``ransomwear''). \cite{43798} describe, among others, the underground market of scareware and ransomware, by showing that (prudent) statistics for CryptoLocker account for \$3 million revenue in 2013-2014. In \cite{Rieck:2008:LCM:1428322.1428330} the authors propose a new method to learn and discriminate malware behaviour. Similarly, in \cite{Rieck:2011:AAM:2011216.2011217} the authors propose a framework to automatically analyse malware behaviour through ML. In particular, the framework allows researchers to identify novel classes of malware having similar behaviour, and to assign labels to the new discovered classes.
In contrast, in \ourname{} we have shown the importance of feature selection to reduce the complexity of the ML algorithms without affecting the performance. This contrasts with some solutions proposed in where no feature selection algorithm is proposed \cite{kolter2006learning}, \cite{Rieck:2011:AAM:2011216.2011217}. In \cite{dimva_paper} the authors analyse 15 different ransomware families, by showing that the large majority of samples implement locking or encrypting techniques through na{\"i}ve techniques. The authors describe simple techniques to detect and stop ransomware, such as monitoring abnormal file activities. This analysis is confirmed by \ourname{} results: by using a limited set features we were able to detect more than 96\% of ransomware. HelDroid \cite{Andronio2015} is a system for recognizing Android ransomware using their typical characteristic, such as functions to lock screen. The detection results for HelDroid are better than those of \ourname{}, due to the fact that the sophistication, and variance, of ransomware for Android is quite limited with respect to that for Windows. Finally, \cite{CryptoLock} is a early-warning detection system for ransomware that checks for file activities and alerts the user in case of suspicious activities, using an union of three features (file type changes, similarity measurement and entropy).

%---conclusion----%
\section{Conclusion} \label{sec:conclusion}
Given the monetary gains achievable, ransomware has become the focus of many cyber-criminals leading to its rapid evolution, and to sophisticated samples able to evade signature-based AVs.  Coping with both new variants of known families, and new families, is therefore of essence. We have shown that ML is a viable and effective approach to detect new variants and families of ransomware for subsequent analysis and signature extraction, and as a complement for AV. Mutual Information has shown to be an effective way of automatically selecting the features, while Regularized Logistic Regression has shown to be an accurate algorithm, easy to train and update, and fast. In terms of results, it compares well with more sophisticated algorithms and leads to much better results than more na\"{\i}ve approaches.

\bibliographystyle{abbrv}
\bibliography{ransomware}  % sigproc.bib is the name of the Bibliography in this case

%\appendix
%\input{appendix}
\balancecolumns

\end{document}